\documentclass[journal]{IEEEtran}

\usepackage{mathtools}
\usepackage{tikz}

\usepackage{color}

\begin{document}
%
\title{Electronic DC-SQUID Emulator}
%
%
%

\author{Josiah~Cochran, C.~S.~Guzm\'{a}n~IV, Eric~Stiers, and~Irinel~Chiorescu%
\thanks{Josiah Cochran is a graduate student with Department of Physics and The National High Magnetic Field Laboratory, Florida State University, Tallahassee, Florida 32310, USA. e-mail: cochran@magnet.fsu.edu}
\thanks{C.~S.~Guzm\'{a}n~IV is a graduate student with Department of Physics and The National High Magnetic Field Laboratory, Florida State University, Tallahassee, Florida 32310, USA. e-mail: csguzman@fsu.edu}
\thanks{Eric Stiers is an electronics engineer with The National High Magnetic Field Laboratory, Florida State University, Tallahassee, Florida 32310. e-mail: stiers@magnet.fsu.edu }
\thanks{Irinel Chiorescu is a professor with Department of Physics and The National High Magnetic Field Laboratory, Florida State University, Tallahassee, Florida 32310, USA e-mail: ic@magnet.fsu.edu}}
\maketitle

\begin{abstract}
Pulsed readout of Direct Current (DC) SUperconducting Quantum Interference Device (SQUID) is crucial for experiments which need to be performed at millikelvin temperatures, such as the readout of superconducting and electron spin based qubits. Pulsed readout algorithms used in these experiments are usually specific to the experimental setup and require some optimization. We present a circuit that emulates the behavior of a DC-SQUID in order to allow the development and evaluation of pulsed readout algorithms at room temperature without the need of a running dilution refrigerator. This novel circuit also constitutes a low cost device which can be used to teach the principles of a DC-SQUID in courses aimed at training the next generation of quantum engineers.
\end{abstract}

\IEEEpeerreviewmaketitle

\section{Introduction}

\IEEEPARstart{W}{e} present a novel circuit design for a DC-SQUID emulator in which all the SQUID parameters are easily tuneable and the result of a DC-SQUID measurement is nearly indistinguishable from its microscopic superconducting counterpart. The presented circuit  emulates DC and pulsed SQUID current-voltage (IV) measurements in the presence of a modulating magnetic flux. Analog computers have been used to calculate the Radio Frequency (RF) response of SQUIDs~ \cite{russer1972influence,hamilton1973analog,werthamer1967analog}, albeit on time scales much slower than that of SQUIDs, but to our knowledge no circuit has been built to directly replace a SQUID in an electronics setup. Pulsed measurements of DC-SQUIDs are required for quantum experiments which need to be performed at millikelvin temperatures, such as readout of superconducting and electron spin based qubits~\cite{chiorescu2003coherent,yue2017sensitive,wernsdorfer2009micro}. Wet dilution refrigerators, which have much higher cooling power due to the use of a liquid helium bath, have the downside of consuming many liters of liquid helium while dry systems can take days to cool down initially and overnight to cool a new sample~\cite{batey2015principles}. The circuit presented in this paper allows for the development and evaluation of DC-SQUID setups with no helium cost and no cool-down time. Furthermore, this circuit can also be used as a training bed for new algorithms or for teaching a course on quantum technologies~\cite{sewani2020coherent,aiello2021achieving,asfaw2021building,lamata2021modernizing}. The circuit presented in this paper was used for the development of a novel feedback algorithm recently demonstrated~\cite{cochran2020dual}.

A DC-SQUID consists of two Josephson junctions (JJ) connected in parallel to create a loop. These junctions are governed by the Josephson equations~\cite{van1981principles}: 
\begin{equation}
	\label{josephson_current}
	I_{jj} = I_c\sin{\phi}
\end{equation}
and 
\begin{equation}
	\label{josephson_voltage}
	V_{jj} = \frac{\Phi_o}{2\pi}\frac{\partial \phi}{\partial t},
\end{equation}
where $I_c$ is the critical current and $\phi$ is the phase difference of the superconducting wave function across the junction, $I_{jj}$ is the current through the junction, $V_{jj}$ is the voltage across the junction. $\Phi_o=h/2e$ is the flux quantum where h is the Planck's constant and $e$ is the electron charge. These two equations describe the super-current that flows when the JJ is in the superconducting state; when $I_c$ is exceeded, the JJ acts as a normal resistor. When two ideal JJ's are placed in parallel to create a DC-SQUID, the current at which the device switches to the resistive state follows the equation~\cite{clarke2006squid}: 
\begin{equation}
	\label{Isw}
	I_{sw} = 2I_c\vert\cos{\pi \frac{\Phi_{ext}}{\Phi_o}}\vert,
\end{equation}
where $I_{sw}$ is the switching current of the SQUID, $I_c$ is the critical current of a single junction, $\Phi_{ext}$ is the externally applied flux to the loop. Based on this relation, the switching current of the SQUID can be used to make an extremely sensitive flux magnetometer~\cite{robbes2006highly,vasyukov2013scanning}.

\begin{figure}[!t]
	\centering
	\includegraphics[width=\columnwidth]{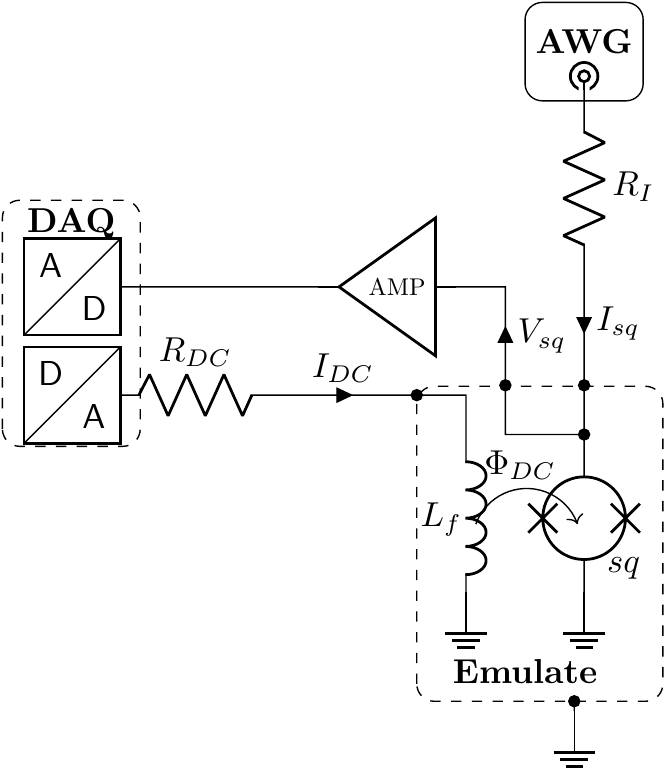}
	\caption{SQUID flux feedback readout. An arbitrary waveform generator (AWG) is used to send a voltage pulse to the SQUID. A series resistance ($R_I$) is added to set the current sent to the SQUID ($I_{sq}$). The SQUID is a short before switching and has a very low impedance in the resistive state so the voltage pulse from the AWG can be considered a current pulse. A sensitive amplifier picks up any small voltage that develops across the SQUID ($V_{sq}$) and detects a switching event as a finite pulse in $V_{sq}$. The voltage output of the D/A converter passes through a series resistor ($R_{DC}$) to apply a current ($I_{DC}$) to an inductor coupling a flux ($\Phi_{DC}$) into the SQUID, with the purpose of creating a flux feedback mechanism (see text). Our circuit emulates the components contained in the dashed box labeled "Emulate".}
	\label{flux feedback method}
\end{figure}

A circuit diagram showing a typical SQUID measurement is given in Fig.~\ref{flux feedback method} (also used in~\cite{cochran2020dual}). An arbitrary waveform generator (AWG) sends a current pulse ($I_{sq}$) to the SQUID ($sq$) while an amplifier picks up the voltage ($V_{sq}$) across the SQUID. If the current pulse is below a switching current ($I_{sw}$), there is no voltage pulse and if the current is above, the SQUID becomes resistive and a voltage pulse is detected on the $V_{sq}$ line. The $I_{sw}$ value is probabilistic due to thermal and quantum fluctuations and a number of $N$ pulses is sent to determine the probability of switching $P_{sw}$ at a certain $I_{sq}$ as the ratio between the number of events showing a $V_{sq}$ pulse and $N$. The current that gives $P_{sw}=50\%$ is used as a measure of $I_{sw}$ for a fixed external flux  ~\cite{chiorescu2003coherent,yue2017sensitive}. One of the methods to perform a pulsed DC-SQUID readout uses a fixed working point (fixed flux such as $P_{sw}=50\%$) by means of a feedback loop mechanism. The algorithm changes the current $I_{DC}$ through an inductively coupled flux bias coil~($L_f$ in Fig.~\ref{flux feedback method}) to keep the SQUID flux constant; the amount of change in $I_{DC}$ is a representation of the change in external flux and thus it constitutes a sensitive flux detection. 

Current pulses $I_{sq}$ can be used to record a pulsed IV curve. For each $I_{sq}$ value one sends multiple current pulses of height $I_{sq}$ and one averages the corresponding SQUID voltages to obtain a current-voltage plot as given in Fig.~\ref{DC_IV} (dashed lines).

\section{SQUID Emulator Circuit Design}

\begin{figure}
	\centering
	\includegraphics[width=\columnwidth]{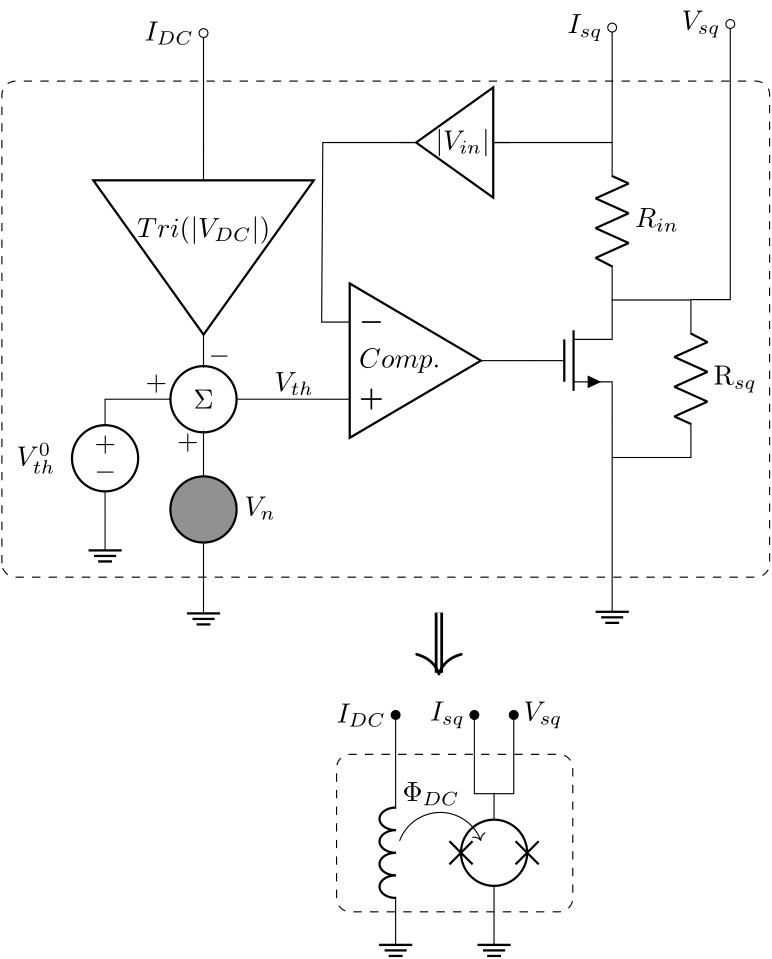}
	\caption{Block Diagram of SQUID Emulator with flux modulation and noise. A SQUID current $I_{sq}$ and DC flux bias current $I_{DC}$ are fed in, and the SQUID voltage is read out. A triangle(x) block (see Fig.~\ref{triangle}) converts $I_{DC}$ to a periodic triangular wave which is subtracted from $V^0_{th}$. A noise source $V_n$ is added to simulate thermal noise. The rest of the block emulates a DC-SQUID at zero flux (see text).}
	\label{block_diagram}
\end{figure}

Based on the SQUID operational principles and the typical measurement scheme outlined in the previous section, the constraints for a circuit that emulates a DC-SQUID can be inferred. The SQUID emulator must have zero resistance up to a certain current $I_{sw}$ at which the SQUID switches to the normal state characterized by a finite resistance $R_{sq}$. The value of $I_{sw}$ must be modulated in a periodic fashion by an external bias current $I_{DC}$ inducing a bias flux ($\Phi_{DC}$ in Fig.~\ref{flux feedback method}).


The circuit for emulating the IV characteristics of a DC-SQUID is shown in Fig.~\ref{block_diagram}. A comparator (Comp.) is used to determine if the SQUID bias is above a switching threshold. A resistor $R_{in}$ converts $I_{sq}$ into a voltage $V_{in}=I_{sq}R_{in}$. If $V_{in}<V_{th}$, the output of the comparator is high and the N-Channel MOSFET short circuits the SQUID resistance $R_{sq}$ (superconducting SC state). If $V_{in}>V_{th}$, the output of the comparator is low and the MOSFET has a finite resistance $R_M$ in parallel to $R_{sq}$ ($R_M||R_{sq}$ represents the normal state). In Fig.~\ref{block_diagram}, the notation $V_{th}^0$ corresponds to zero flux $\Phi_{DC}=0$, while $V_{th}<V_{th}^0$ would indicate the presence of a flux. A voltage source tunes $V_{th}^0$, emulating the desired value of the switching current at zero flux. An active rectifier ~\cite{sedra2010} delivers $\vert V_{in} \vert$, and therefore a current less than $-I_{sw}$ also causes a switching event since a SQUID IV curve is antisymmetric in current. As long as the SQUID voltage $V_{sq}$ is small, the MOSFET is not brought into saturation and the emulator works well for positive and negative current pulses. In a typical DC IV measurement $I_{sq}$ is ramped up above $I_{sw}=V_{th}/R_{in}$ and then it is ramped down. The SQUID will switch back to the SC state at the retrapping current $I_r$ given by: 
\begin{equation}
	\label{retrap_current}
	I_{r} = V_{th}/(R_{in}+R_{sq}||R_M).
\end{equation} 
Since $R_{sq}\neq0$, $I_r<I_{sw}$ which correctly emulates a hysteresis in the IV curve as shown in Fig.~\ref{DC_IV}.

In order to emulate the flux modulation and the probabilistic nature of the switching as influenced by thermal fluctuations, two addition are needed: triangular modulation $Tri(|V_{DC}|)$ and noise source $V_n$. The first unit to be introduced, is the triangular modulator that takes a linearly increasing voltage, such as $V_{DC}$, and transforms it into a triangular wave which is used to emulate the modulation behavior of $I_{sw}$ for SQUIDs featuring Dayem bridges~\cite{yue2017sensitive}.

\begin{figure}
	\centering
	\includegraphics[width=\columnwidth]{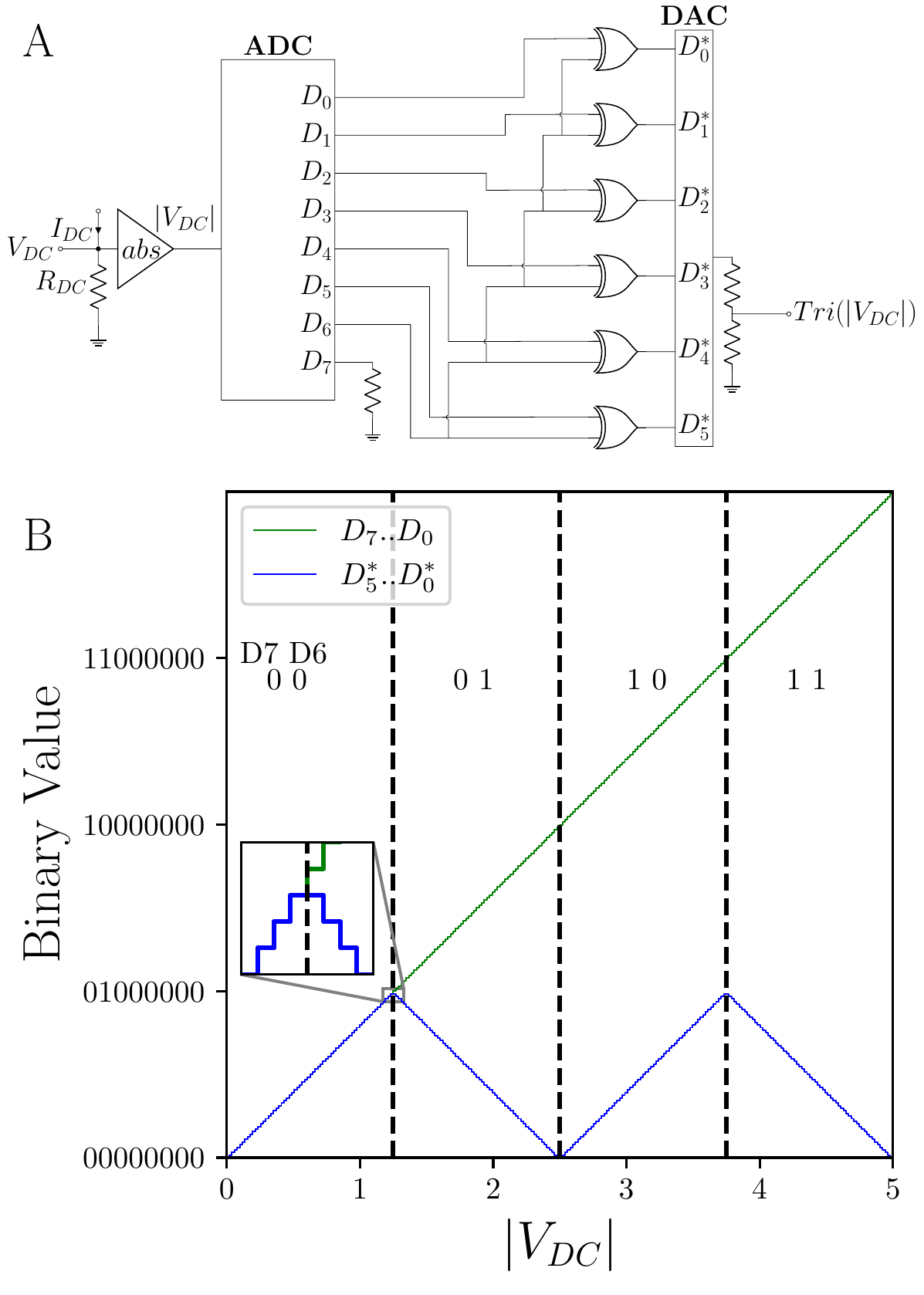}
	\caption{A) Mixed Domain Linear to Triangular Periodic Converter. This circuit contains an active rectifier on the input (abs) to symmetrize the output for positive and negative input voltages. The next stage is a free running ADC, for which an ADC0804 was used, followed by an array of XOR gates which remap the digital values as shown in B). The digital values are then passed to R2R resistor ladder DAC. B) Shows the output digital values $D^*_{5..0}$ (blue curve) and the input digital values $D_{7...0}$ (green curve) versus input voltage $V_{DC}$. For each increment of the 2 bit value of D7 D6, the slope of the output changes.}
	\label{triangle}
\end{figure}

The triangular modulation works in the following way. An analog-digital converter ADC0804 (see Fig.~\ref{triangle}) performs an 8-bit digitization of the modulation voltage $V_{DC}$. The 8 bits are used as follows: (i) the digit $D_7$ indicates the period of the modulation: $V_{DC}<V_{max}/2$ or $V_{DC}>V_{max}/2$; (ii) the digit $D_6$ establishes the slope sign for the output voltage (0 is positive and 1 is negative); (iii) the remaining six bits create an output voltage with $2^6=64$ steps of resolution as indicated by a blue line in Fig.~\ref{triangle}B. This step is done using a R2R ladder~\cite{horowitz1989art} as digital-to-analog converter (DAC). 

All six XOR gates share $D_6$ as one of the inputs and $D_{5\ldots0}$ as the other one, to generate the outputs $D^*_{5\ldots0}$. For $D_6=0$, we have $D_{0\ldots5}^*=D_{0\ldots5}$ and thus the blue and green curves overlap in Fig.~\ref{triangle}B. All six bits start from 0 and count up. When the $V_{DC}$ is a quarter of the maximum voltage, $D_6=1$ and $D_{0\ldots5}^*=\overline{D_{0\ldots5}}$ meaning that all six bits start from 1 and count down. This ensure the continuity of the 6-bit output binary value of the blue curve when going through a maximum. At $V_{DC}=V_{max}/2$, $D_{6..0}=0$ and $D_7=1$. Once again, all six bits start from 0 and count up and the second cycle runs until $V_{DC}=V_{max}$. More cycles can be obtained with additional bits for the ADC but for the purposes of emulating a SQUID behavior, four cycles are sufficient (two cycles for each polarity of $V_{DC}$). The final output voltage will show four periods of triangular modulation with a resolution of $2^{-6}$ of the maximum voltage (see steps in the inset of Fig.~\ref{triangle}B). The final stage of the modulator is a voltage divider meant to scale down the modulation amplitude from Volts to tens of mV, to emulate a typical SQUID.

In the circuit of Fig.~\ref{block_diagram}, one notes that $V_{th}^0$ is summed with a noise source $V_n$ and the modulated voltage $-Tri(|V_{DC}|)$, obtained as explained above, giving $V_{th}=V_{th}^0+V_n-Tri(|V_{DC}|)$. The noise source simulates thermal noise on the switching process, as described in the next section. The flux modulation $I_{sw}(\Phi)$ is emulated by the periodical behavior of $V_{th}(I_{DC})$. Note that for $V_{th}^0$ tuned to be larger than the modulation depth of $Tri(|V_{DC}|)$, $V_{th}$ has a positive value even in the minimum of the modulation.  This is important to properly emulate SQUIDs with Dayem bridges which have an important kinetic inductance in the loop and $I_{sw}(\Phi_{DC})$ does not reach zero at half quantum flux $\Phi_0/2$.

If a sinusoidal modulation of $I_{sw}$ is desired instead of the linear branches of a triangular modulation, a circuit different from that of Fig.~\ref{triangle} is needed. For instance, a function $\cos(x)$ can be created by using analog multipliers such as the AD633. A series of analog multipliers in conjunction with positive and negative gain amplifiers can approximate the Taylor expansion of $\cos(x)=\sum_0^N (-1)^k\frac{x^{2k}}{(2k)!}$ up to a desired order $N$.  The sinusoidal modulation is desirable when trying to model a tunnel junction. This type of modulation is not physically implemented here, but the concept is quite simple and similar to designs of old analog computational circuits~\cite{russer1972influence}.
  

The experimental noise is modeled by summing a noise voltage, $V_n$ in Fig.~\ref{block_diagram}, with $V_{th}^0$. A Siglent SDG1032X was used to create white noise $V_n$. A reverse biased Zener diode AC coupled to an operational amplifier with a feedback potentiometer for noise level control could also be used, but it was found that any non-linearity combined with the needed large amplification makes the circuit prone to oscillations and thus to resonances in the noise spectra. A synthesized noise source gives a much more reliably flat power spectral density. The added noise to the threshold means that the flux feedback method can be tested for stability at different levels of noise present in the SQUID circuitry. 

\section{SQUID Emulator Results}

In the following sections, experimental measurements are performed on the SQUID emulator which are common diagnostic measurements for determining flux sensitivity and SQUID parameters. 

\begin{figure}[!t]
	\centering
	\includegraphics[width=\columnwidth]{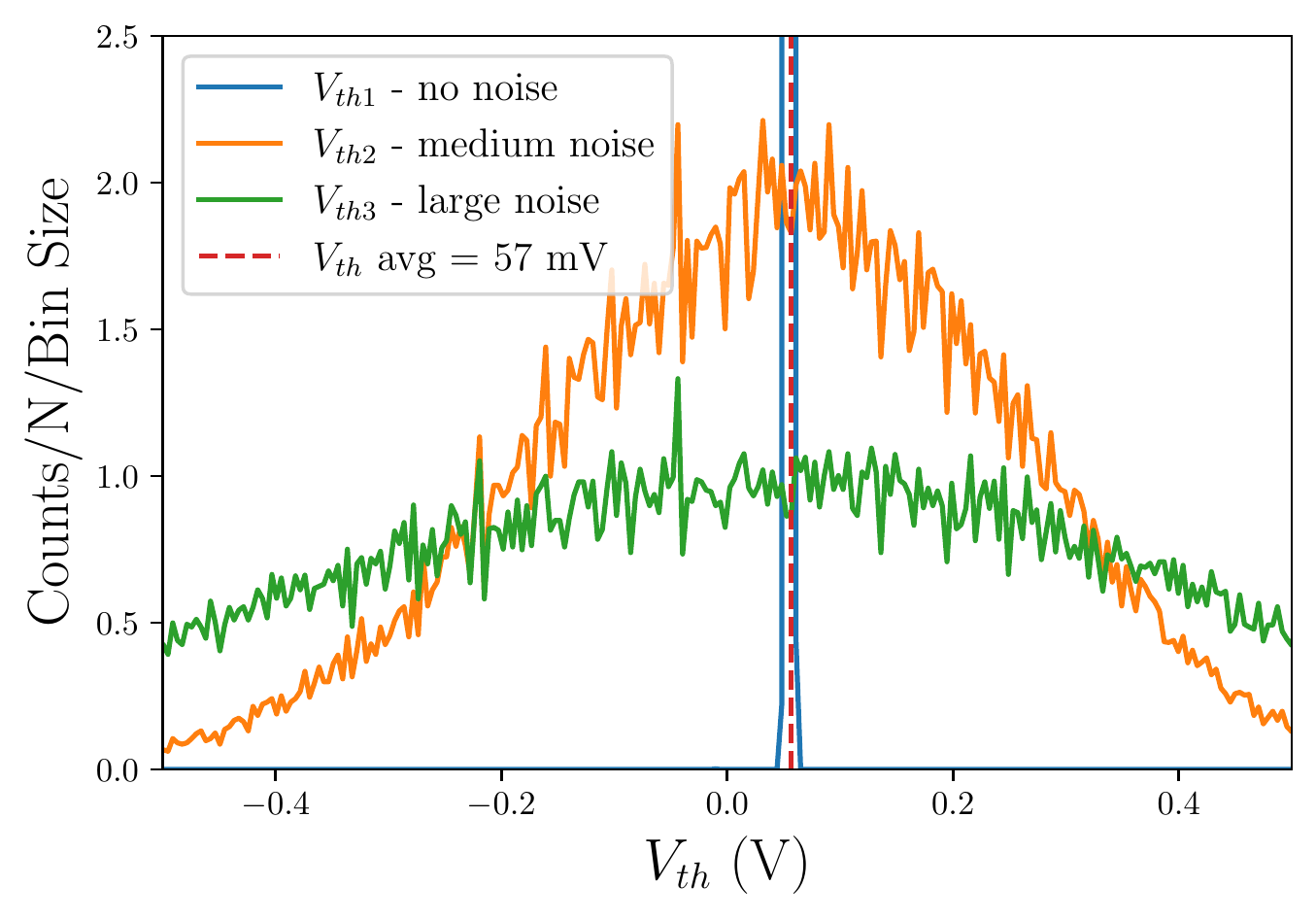}
	\caption{Histograms of $V_{th}$ for no, medium, and large noise. $V_{th}=V_{th}^0+V_n-Tri(|V_{DC}|)$. For large noise, standard deviation of $V_n$ is chosen to be such that the hysteresis of the DC IV curve just closes (see Fig.~\ref{DC_IV}). Medium noise standard deviation of $V_n$ is chosen to be half that value. No noise shows the intrinsic noise of the circuit.}
	\label{noise_hist}
\end{figure}
\begin{figure}[!t]
	\centering
	\includegraphics[width=\columnwidth]{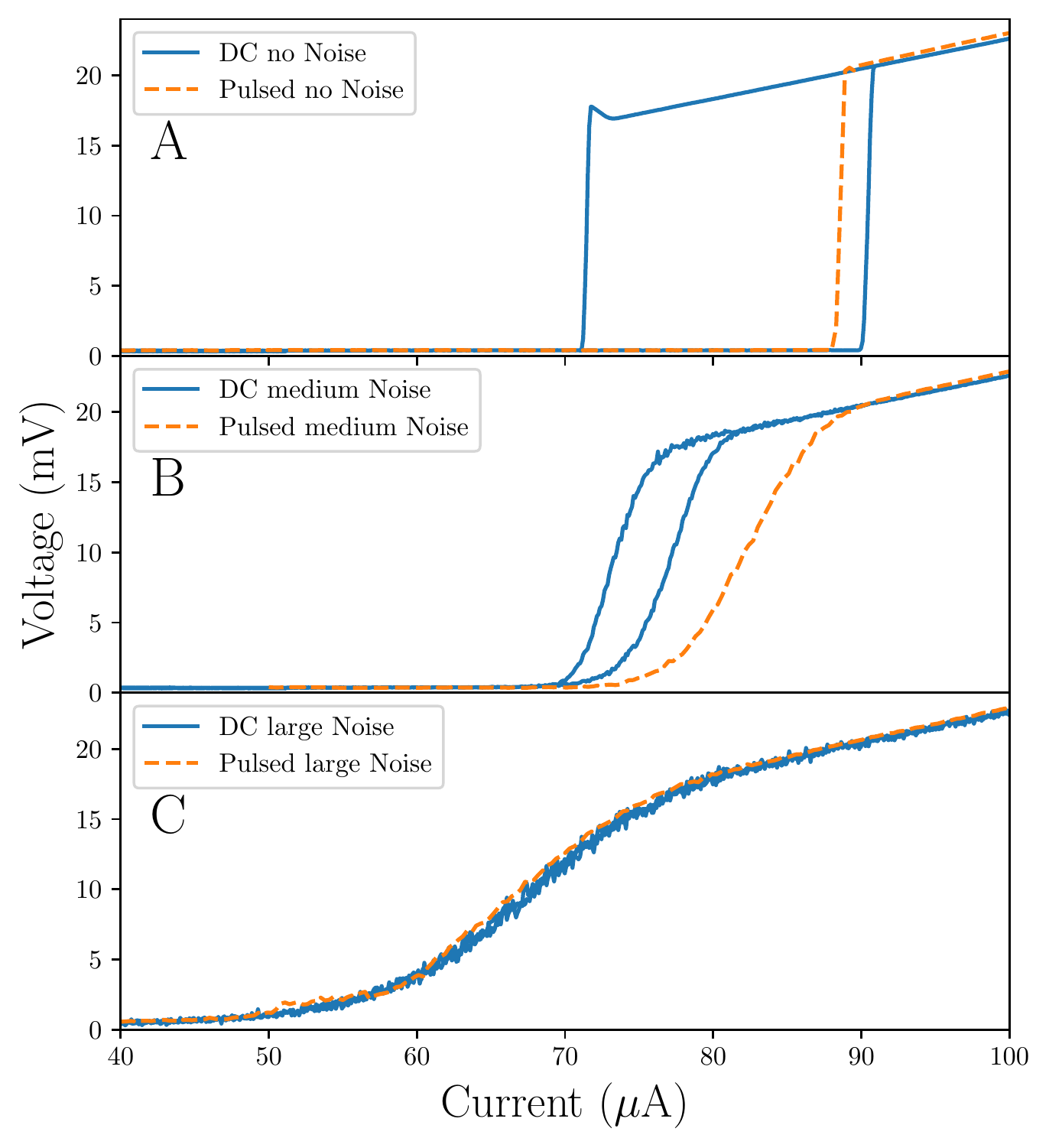}
	\caption{ Pulsed and DC IV curves. A) No added noise. DC IV displays a wide hysteresis while pulsed curve switches at a single current close to the DC switching current on the rising current edge. B) Medium noise. DC IV curve hysteresis narrows with an overall lowering and broadening of the switching current. C) Large noise. The Pulsed and DC curves become nearly indistinguishable. The hysteresis in the DC curve has vanished.}
	\label{DC_IV}
\end{figure}

The experiments in the following sections were executed with varying degrees of noise to see how well the emulator is able to reproduce a SQUID-like behavior in the presence of noise. Fig.~\ref{noise_hist} shows histograms of $V_{th}$ for the different levels of noise used throughout the experiment. These curves will be helpful in understanding how the switching of the SQUID  is affected by noise. $V_{th}$ was measured with a data acquisition system (Adwin Gold) by connecting a probe in the circuit of Fig.~\ref{block_diagram} at the comparator non-inverting input and then sampling voltage values to create a histogram. The zero noise $V_{th}$ histogram shows the intrinsic noise of the emulator. The large noise setting is such that the hysteresis in the IV curve (see Fig.~\ref{DC_IV}C) just vanishes and the medium noise is set at half that level (that is, a Gaussian with half the width of that for the large noise case). The histograms show $V_{th}$ for these three values of $V_n$ with $I_{DC}=0$ therefore the mean of all histograms is $V_{th}^0$ and the standard deviation is the standard deviation of $V_n$. $R_{in}$ for this device was 1~k$\Omega$ so $\mu$A in $I_{sw}$ translates to mV in $V_{th}$.

The DC and pulsed IV curves for the three $V_{th}$ noise levels of Fig.~\ref{noise_hist} are shown in Fig.~\ref{DC_IV}. For a single DC IV curve, measured by ramping $I_{sq}$ up and down, the switching event creates a very sharp voltage pulse, but the value of $I_{sw}$ is probabilistic due to noise. For this reason, $N=5000$ curves are averaged to obtain the DC IV curves shown in Fig.~\ref{DC_IV}. For the smallest noise level (panel A), the hysteretic nature of the emulator is clear and the transition from superconducting to resistive state is very sharp. The averaging effect of the noise leads to a broadening of the switching over a certain probabilistic region of $I_{sw}$ and eventually leads to the disappearance of the hysteresis. In the case of Pulsed IV curves, for each value of $I_{sw}$, $N$ current pulses are sent and the resulting voltages are averaged (there is no retrapping due to the pulsed nature of the measurement). Again, the noise broadens the switching region. As a side note, one observes that $V_{th}$ mean from Fig.~\ref{noise_hist} is not the $\sim$90$\mu$A seen in Fig.~\ref{DC_IV}A, due to an offset between the positive and negative inputs of the AD820 operational amplifier used as a comparator; this aspect is inconsequential to the operation of the emulator. The value of $I_r$ can be calculated using Eq.~\ref{retrap_current}. For $V_{th}=90$~mV,~$R_{in}=1k\Omega$ and $R_{sq}||R_M$ measured to be 225~$\Omega$, one gets  $I_r=73\mu$A, matching the value observed in Fig.~\ref{DC_IV}A.

\begin{figure}[!t]
	\centering
	\includegraphics[width=\columnwidth]{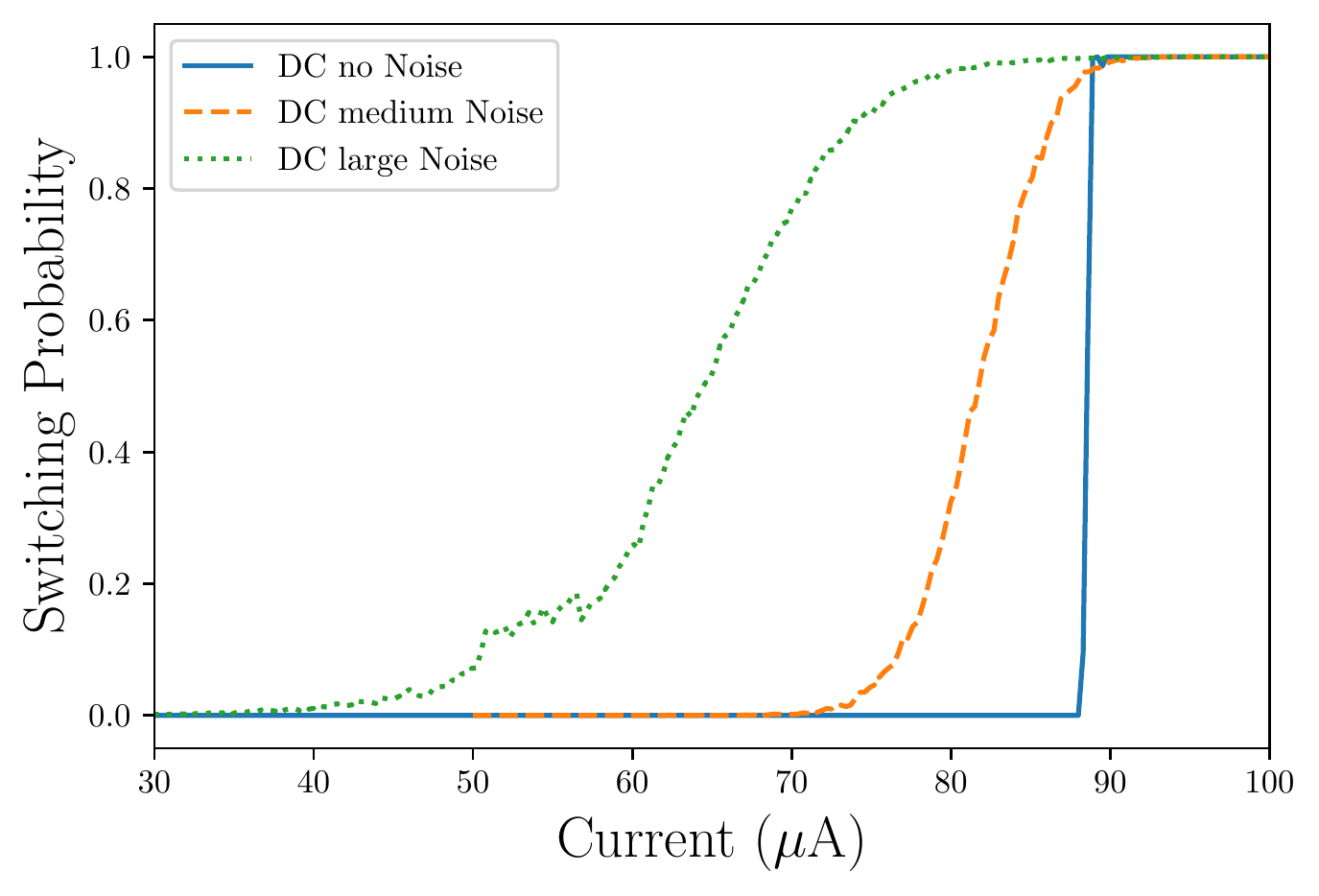}
	\caption{S-Curves for no, medium, and high noise. In the absence of noise, the S-Curve is extremely sharp. Adding noise broadens the S-Curve and lowers the switching current which in turn decreases the absolute flux sensitivity.}
	\label{scurves}
\end{figure}

The cumulative distribution of switching currents $P_{sw}(I_{sw})$, or S-Curve, is an essential diagnostic for determining SQUID sensitivity as a flux to current transducer. The S-Curves of Fig.~\ref{scurves} show that the addition of noise broadens and shifts the switching curve to lower values. This allows a simple mapping of switching probability $P_{sw}$ to switching current and thus flux. The shift of the S-curve to lower currents is a natural consequence of the fluctuations creating a finite probability to switch at lower values of the current, as seen in actual devices.

\begin{figure}[!t]
	\centering
	\includegraphics[width=\columnwidth]{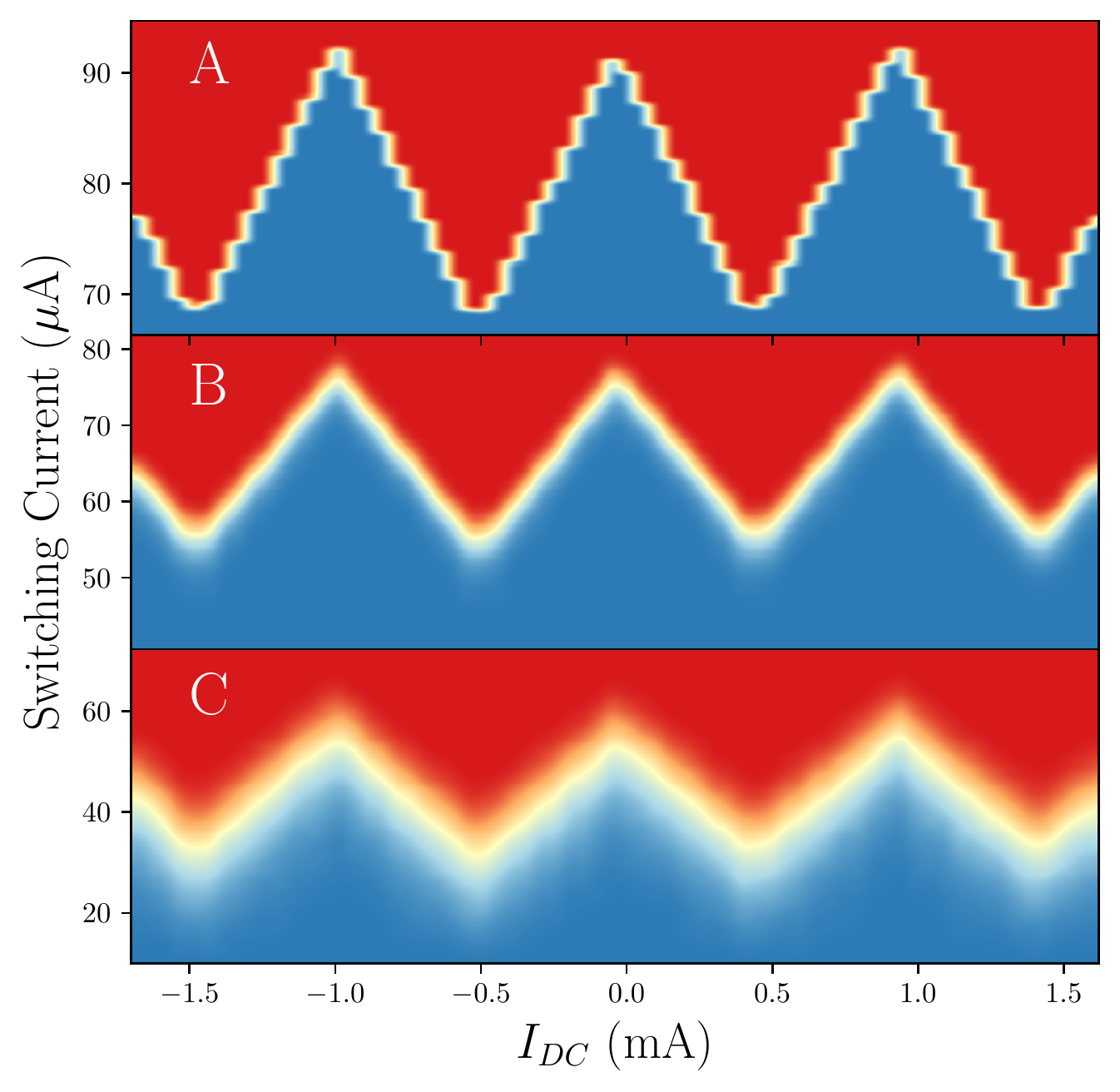}
	\caption{Modulation curves for different levels of noise. Red indicates a $P_{sw}$ of 100\% and blue a $P_{sw}$ of 0\%. A) Modulation curve for intrinsic device noise. The stairstep pattern is due to the discrete steps in $I_{DC}$. B) Modulation curve for medium noise. The vertical width of the modulation is broadened displaying the uncertainty in $P_{sw}$ caused by noise; the average value also shifts down. C) Modulation curve for large noise. The modulation is broadened even further and shift down further. The depth of modulation for all 3 curves is similar and is chosen to not go all the way to zero $I_{sw}$, to model a Dayem bridge device with kinetic inductance in the loop.}
	\label{mod_curves}
\end{figure}
To emulate the flux sensitivity and, if need be, to evaluate the flux feedback needed to maintain the SQUID at a fixed operating point, the $I_{sw}(\Phi_{DC})$ modulation and corresponding statistics must be known. For no noise case, the switching is abrupt and displays a pristine modulation curve as shown in Fig~\ref{mod_curves}A. The stair-stepping pattern is due to the discrete nature of the steps taken in $I_{DC}$. Increasing the noise decreases $I_{sw}$ and blurs the modulation curve (Fig~\ref{mod_curves}B and C). The modulation range of the emulator is set by a sense resistor $R_{DC}$ and the gain of the next operational amplifier (see Fig~\ref{triangle}A) such that $I_{DC}$=2~mA gives $V_{DC}$=5~V, the full scale of the ADC. The modulation of a Nb Dayem bridge SQUID~\cite{yue2017sensitive} is nearly indistinguishable from that shown in Fig.~\ref{mod_curves}.

\section{Conclusion}

An electronic circuit which accurately models the DC switching and modulation characteristics of a DC-SQUID was developed, built, and measured. The emulator is essential to plan and test efficiently SQUID based flux detectors, used in superconducting and spin based quantum computing. At the same time, the circuit is useful to those getting started in the growing field of quantum electronics; the emulator can be used to teach laboratories in the emerging field of quantum engineering which has a growing demand for trained work force. Results on a real SQUID of the flux feedback method developed using this emulator design are presented in~\cite{cochran2020dual}.
\appendices

\section*{Acknowledgment}

This work was performed at NHMFL at the Florida State University and supported by the National Science Foundation through NSF/DMR-1644779 and the State of Florida.

\ifCLASSOPTIONcaptionsoff
  \newpage
\fi

\bibliographystyle{IEEEtran}
\bibliography{squid_emulator}

\end{document}